\documentclass[graybox,natbib,nosecnum]{svmult}
\bibpunct{(}{)}{;}{a}{}{,} 

\pdfoutput=1   

\usepackage{mathptmx}       
\usepackage{helvet}         
\usepackage{courier}        
\usepackage{type1cm}        

\usepackage{makeidx}         
\usepackage{graphicx}        
\usepackage{multicol}        
\usepackage[bottom]{footmisc}
\usepackage[normalem]{ulem}	
\usepackage{hyperref}  

\usepackage{soul}   

\def\arcm{\hbox{$^\prime$}}                

\makeindex             

\begin{document}

\title*{SPECULOOS exoplanet search and its prototype on TRAPPIST}
\author{Artem Burdanov, Laetitia Delrez, Micha{\"e}l Gillon, Emmanu{\"e}l Jehin, and the SPECULOOS \& TRAPPIST teams}
\authorrunning{The SPECULOOS \& TRAPPIST teams}
\institute{Artem Burdanov \at Space sciences, Technologies and Astrophysics Research (STAR) Institute, Universit{\'e} de Li{\`e}ge, All{\'e}e du 6 Ao{\^u}t 17, 4000 Li{\`e}ge, Belgium, \email{artem.burdanov@ulg.ac.be}
\and Laetitia Delrez \at Astrophysics Group, Cavendish Laboratory, J.J. Thomson Avenue, Cambridge CB3 0HE, UK, \hbox{\email{lcd44@cam.ac.uk}}
\and Micha{\"e}l Gillon \at Space sciences, Technologies and Astrophysics Research (STAR) Institute, Universit{\'e} de Li{\`e}ge, All{\'e}e du 6 Ao{\^u}t 17, 4000 Li{\`e}ge, Belgium, \email{michael.gillon@ulg.ac.be}
\and Emmanu{\"e}l Jehin \at Space sciences, Technologies and Astrophysics Research (STAR) Institute, Universit{\'e} de Li{\`e}ge, All{\'e}e du 6 Ao{\^u}t 17, 4000 Li{\`e}ge, Belgium, \email{ejehin@ulg.ac.be}}
%
%
\maketitle

\abstract{One of the most significant goals of modern science is establishing whether life exists around other suns. The most direct path towards its achievement is the detection and atmospheric characterization of terrestrial exoplanets with potentially habitable surface conditions. The nearest ultracool dwarfs (UCDs), i.e. very-low-mass stars and brown dwarfs with effective temperatures lower than \hbox{2700 K}, represent a unique opportunity to reach this goal within the next decade. The potential of the transit method for detecting potentially habitable Earth-sized planets around these objects is drastically increased compared to Earth-Sun analogs. Furthermore, only a terrestrial planet transiting a nearby UCD would be amenable for a thorough atmospheric characterization, including the search for possible biosignatures, with near-future facilities such as the James Webb Space Telescope. In this chapter, we first describe the physical properties of UCDs as well as the unique potential they offer for the detection of potentially habitable Earth-sized planets suitable for atmospheric characterization. Then, we present the SPECULOOS ground-based transit survey, that will search for Earth-sized planets transiting the nearest UCDs, as well as its prototype survey on the TRAPPIST telescopes. We conclude by discussing the prospects offered by the recent detection by this prototype survey of a system of seven temperate Earth-sized planets transiting a nearby UCD, TRAPPIST-1.} 


\section{Introduction}
Confined for centuries to the rank of a pure speculation, the existence of life outside our Solar System is now at the edge of gaining its status of a testable scientific hypothesis. Since the first discoveries of planets orbiting other stars than the Sun \citep{1992Natur.355..145W,1995Natur.378..355M}, more than three thousand of such exoplanets have been detected at an ever increasing rate \citep{2011A&A...532A..79S,2014PASP..126..827H}. The present exoplanets harvest is not only composed of gas and ice giants, but also includes a steeply growing fraction of small, potentially terrestrial planets. In parallel to this galore of detections, many projects aiming to characterize exoplanets have reached success in the last decade, bringing notably first pieces of information on the atmospheric properties of giant exoplanets. Nearly all of these atmospheric studies have been made possible by the transiting configuration of the probed planets. Indeed, the special geometrical configuration of transiting planets offers the detailed study of their atmosphere without the cost of spatially resolving them from their host stars \citep{2010arXiv1001.2010W}. The first atmospheric studies of transiting "hot Jupiters" performed with space- and ground-based instruments have provided initial glimpses at the atmospheric chemical composition, vertical pressure-temperature profiles, albedos, and circulation patterns of extrasolar worlds \citep{2017arXiv170100493D}. On paper, exporting the techniques developed for these pioneering first studies of transiting gas giants to the atmospheric characterization of terrestrial planets orbiting in the habitable zone (HZ, \citealt{2013ApJ...765..131K}) of their host star looks like a promising path to search for life outside our Solar System in the near future. The relevance of this approach relies on the discovery of suitable transiting planets, i.e. HZ terrestrial planets transiting a host star bright and small enough to lead to adequate signal-to-noise ratios (SNRs) for spectroscopic detection of biosignatures \citep{2016AsBio..16..465S}, assuming realistic observational programs with the upcoming astronomical facilities. Most studies in this domain have focused on the James Webb Space Telescope (JWST) \citep{2009ASSP...10..123S, 2009ApJ...698..519K, 2011A&A...525A..83B, 2013Sci...342.1473D, 2016MNRAS.461L..92B}, because its orbit, large aperture and infrared (IR) sensitivity make it \textit{a priori}  the most promising facility for such atmospheric characterizations. All these studies agree on the fact that the best suitable target for biosignatures detection would be an habitable terrestrial planet transiting one of the nearest ultracool dwarfs (UCDs). Indeed, UCDs are so small and faint that they do not drown out the signals of Earth-sized exoplanets, allowing us to detect and study in great depth such small planets.

\section{What are ultracool dwarfs?}

UCDs are traditionally defined as dwarf stars and brown dwarfs having effective temperatures $\mathrm{T_{eff} < 2700~K}$, luminosities  $\mathrm{L \leq 10^{-3}~L_{\odot}}$, and spectral types later than M6, including L, T and Y dwarfs \citep{2005ARA&A..43..195K, 2011ApJ...743...50C}. In these conditions, the atmospheres of UCDs are rich in molecular gases (H$_2$O, CO, TiO, VO, CH$_4$, NH$_3$, CaH, FeH) and condensed refractory species (mineral and metal condensates, salts and ices), producing complex spectral energy distributions, strongly influenced by composition and chemistry, that peak at near- and mid-IR wavelengths.  UCDs have masses below $\mathrm{\approx 0.1~M_{\odot}}$, extending below the hydrogen burning minimum mass (HBMM) of $\mathrm{0.07~M_{\odot}}$, into the realm of non-fusing brown dwarfs \citep{1962AJ.....67S.579K,1963ApJ...137.1121K, 1963PThPh..30..460H}. Supported from gravitational collapse primarily by electron degeneracy pressure, these objects are the most compact and dense hydrogen-rich bodies in the Galaxy, with radii reaching a minimum of $\mathrm{R \approx 0.08-0.10~R_{\odot}}$ near the HBMM and core densities potentially as high as $\mathrm{1000~g/cm^3}$ \citep{1993RvMP...65..301B}. These dense interiors may sample exotic states of matter (e.g., metallic and crystalline hydrogen), and make UCDs fully convective and well-mixed in their interior composition. Convection, coupled with their low fusion rates, implies lifetimes of tens of trillions of years for stellar UCDs, while substellar UCDs persist indefinitely but with ever-decreasing temperatures and luminosities.
 
While few UCDs were known prior to the mid-1990s, the proliferation of red optical and near-IR surveys over the past 20 years has dramatically increased the size of the known population, including recent discoveries of some of the nearest systems to the Sun: the L dwarf plus T dwarf binary Luhman 16AB at 2.0~pc \citep{2013ApJ...767L...1L}; the Y dwarf WISE 0855-0714 with $\mathrm{T_{eff}~\approx~250~K}$ at $\mathrm{2.3~pc}$ \citep{2014ApJ...786L..18L}; and the M dwarf plus T dwarf binary WISE 0720-0846, which passed within $\mathrm{50,000~AU}$ of the Sun in the past 100,000~years \citep{2015AJ....149..104B, 2015ApJ...800L..17M}. Overall, UCDs appear to be less numerous than more massive M stars ($\mathrm{0.1~M_{\odot} < mass < 0.5~M_{\odot}}$), but are more abundant than solar-type FGK stars in the immediate solar neighbourhood (e.g., $>5:1$ UCDs:G~dwarfs in the $\mathrm{8~pc}$ sample; \citealt{2012ApJ...753..156K}).

There are a number of unique characteristics of UCDs that are relevant to understanding exoplanet companions that warrant mention. First, their low $\mathrm{T_{eff}}$ reduce the coupling of photospheric gas to internally-generated magnetic fields, resulting in a general decline in the relative strength and incidence of optical (H-alpha, Ca II) and X-ray nonthermal emission, particularly among the L and T (and presumably Y) dwarfs \citep{2000AJ....120.1085G, 2006ApJ...648..629B, 2016ApJ...826...73P} . This makes the immediate environment around UCDs somewhat more benign than active M dwarfs. Nevertheless, flaring emission (in some cases dramatic, \citealt{2014ApJ...781L..24S}) persists in these objects, as does nonthermal radio emission, both indicating the presence of strong magnetic fields. The reduction in magnetic coupling reduces angular momentum loss, so that UCDs are generally rapidly rotating bodies with periods as short as 1-2 hours and rotational $\mathrm{v}sin\mathrm{i}$ measurements as high as $\mathrm{80~km/s}$ \citep{2010ApJ...723..684B, 2015ApJ...799..154M}. 

\section{Ultracool dwarfs and planets}

Despite UCDs represent a significant fraction of the Galactic population, their planetary population is still a nearly uncharted territory. As of today, only nine {\it bona fide} planets have been found in orbit around UCDs, the seven transiting Earth-sized planets of TRAPPIST-1 (\citealt{2017Natur.542..456G}, see below), the $\mathrm{\sim3~M_{E}}$ planet MOA-2007-BLG-192Lb and $\mathrm{\sim1.3~M_{E}}$ planet OGLE-2016-BLG-1195Lb detected by microlensing around a distant UCD stars (\citealt{2012A&A...540A..78K,2017arXiv170308548S}). This microlensing planets are very interesting, because it demonstrates that UCDs can form planets more massive than the Earth, despite the low mass of their protoplanetary discs. On their side, the Earth-sized planets transiting TRAPPIST-1 indicate that compact systems of small terrestrial planets are probably common around UCDs, as TRAPPIST-1 is one of only $\sim50$ UCDs targeted by the SPECULOOS prototype survey ongoing on the TRAPPIST-South telescope since 2011 (see below). First limits on the occurrence rate of short-period planets orbiting brown dwarfs were reported by \cite{2017MNRAS.464.2687H}, who found that within a 1.28~d orbit, the occurrence rate of planets with a radius between 0.75 and 3.25~$\mathrm{R_{\oplus}}$ is lower than $67\pm1\%$.

These planet detections are consistent with the growing observational evidence that young UCDs are commonly surrounded by protoplanetary discs (e.g., \citealt{2007prpl.conf..443L}) which, while containing less mass, appear to persist longer as compared to solar-type stars \citep{2012ARA&A..50...65L}. Furthermore, young UCDs also exhibit the hallmarks of pre-planetary formation: evidence of disc accretion, circumstellar disc excess, accretion jets, and planetesimal formation (e.g., \citealt{2000ApJ...545L.141M, 2003ApJ...593L..57K, 2005Natur.435..652W, 2011ASPC..448..469P, 2013ApJ...764L..27R}). 

Nearly unconstrained by direct observations for objects below  $\mathrm{\sim0.2~M_{\odot}}$, planetary formation models agree on the fact that UCDs should be able to form mostly terrestrial planets \citep{2007MNRAS.381.1597P}, but disagree on their typical mass and chemical composition. For instance, \cite{2007ApJ...669..606R} predicted systems of short-period inhospitable metal-rich terrestrial planets that rarely exceed the mass of Mars, while \cite{2009Icar..202....1M} predicted systems of more massive volatile-rich planets, that should be better suited for the emergence of life. \cite{2017A&A...598L...5A} predict Earth-sized planets, that are volatile rich if protoplanetary discs orbiting low-mass stars are long lived. 

When combined with the observational evidence that short-period low-mass planets are common around solar-type stars and tend to be found in nearly coplanar closely packed multiplanetary systems (e.g. \citealt{2012A&A...541A.139F,2016ApJ...816...66B,2012ApJ...747..144M,2015ApJ...801...18M}), and with the discovery of the TRAPPIST-1 system, all these considerations lead to the expectation that the typical planetary system around UCDs should be reminiscent to the Jovian system, with (water-rich or not) terrestrial planets replacing the Galilean moons. If this prediction is valid, most UCDs should have terrestrial planets within or close to their HZ. Indeed, due to their small sizes and low temperatures, the HZs of UCDs are located very close by, at just a few percent of one au \citep{2011A&A...535A..94B, 2013ApJ...765..131K}, or even less for brown dwarfs. If systems of terrestrial planets were confirmed to be common around UCDs, then the archetypical terrestrial planet in our Galaxy would not be Venus, the Earth or Mars but rather a tidally-locked red world like those of TRAPPIST-1.


\section{SPECULOOS: seizing the UCDs opportunity}

Because of the low luminosities and small sizes of UCDs, and the resulting large planet-to-star flux and size ratios, expected SNRs on the detection of spectroscopic signatures in the atmosphere of a transiting habitable Earth-sized planet are more favorable for UCDs than for any other host. \cite{2009ApJ...698..519K} derived SNR expectations for atmospheric biosignatures measured with JWST for Earth-sized planets transiting putative M0 to M9 dwarf stars at $\mathrm{10~pc}$. Scaling these SNRs with the distance to the Earth and considering $\mathrm{SNR=10}$ as an absolute lower limit to constrain properly the atmospheric composition, one can derive the following upper limits on the distance (and on the corresponding J-band magnitude) for the different spectral types of UCD stars: 30~pc ($\mathrm{J=12.6}$) for the M7, 34~pc ($\mathrm{J=13.3}$) for the M8, 40~pc ($\mathrm{J=14.0}$) for the M9 and latter (the spectral type range of UCD stars extends down to $\sim$~L2.5, \citealt{2014AJ....147...94D}). The numbers of UCD stars in the close solar neighbourhood ($<8$~pc) are well known (e.g. \citealt{2008AJ....136.1290R, 2007AJ....133.2825R}), and the corresponding number densities for the different spectral types can be coupled to the distance limits derived above to estimate the number of possible targets as 800 UCD stars. Because of the proximity of the HZ for these UCD stars, the relatively small number of them being bright enough for JWST is balanced by a strongly increased geometric transit probability.  Assuming that each of these 800 nearby UCD stars has a terrestrial planet in its HZ leads indeed to an expected sample of about 20 habitable planets waiting to be studied by JWST and other future facilities, to which one should add the planets orbiting outside the HZ. In addition to these 800 UCD stars, about 200 brown dwarfs are nearby and bright enough to enable the study of the atmospheric composition of short-period transiting Earth-sized planets with JWST (e.g. \citealt {2011A&A...525A..83B}). In total, there are thus about 1000 opportunities in the sky (800 stars + 200 brown dwarfs) to detect Earth-sized planets well suited for detailed atmospheric characterization - including biosignatures \hbox{detection -}  with current and near-future technology. 

From a transit-search perspective, UCDs provide two important observational advantages. First, their small size leads for Earth-sized planets to have transit depths from a few 0.1\% up to $>1\%$, similar to the typical transit depths for the dozens of Jupiter-size planets detected around solar-type stars by wide-field ground-based surveys like WASP (\citealt{2007MNRAS.375..951C}) and HATNet (\citealt{2007ApJ...656..552B}). Secondly, the proximity of their HZ makes the transits of habitable planets have periodicities of a few days similar to gas giants in close orbit around solar-type stars, which translates into a required photometric monitoring of much smaller duration than for \textit{bona fide} Earth-Sun twin systems. This means that a transit search targeting the 1000 nearest  UCDs could be done within a few years with a realistically small number of telescopes, despite that it should monitor each target individually, as nearby UCDs are spread all over the sky. This is the goal and concept of our project SPECULOOS (\textbf{S}earch for habitable \textbf{P}lanets \textbf{EC}lipsing \textbf{UL}tra-c\textbf{OO}l \textbf{S}tars). The project is led by the  University of Li\`ege (Belgium) in collaboration with the Cavendish Laboratory of the University of Cambridge (UK) and the King Abdulaziz University (Saudi Arabia).

\section{The SPECULOOS Southern Observatory}

\begin{figure}
\centering
\includegraphics[scale=.45]{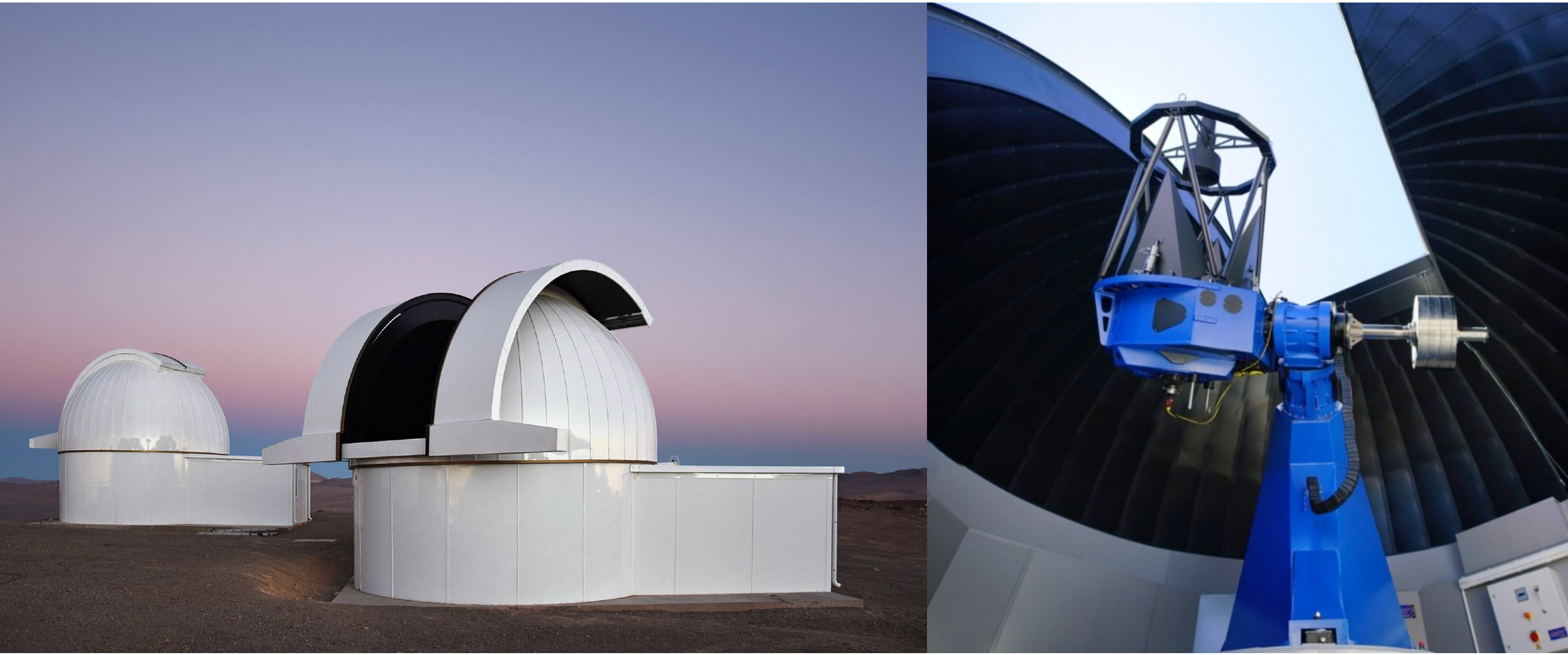}
\caption{\textbf{Left.} First two domes of the SPECULOOS Southern Observatory at ESO Paranal Observatory (Chile). Credit: ESO/G. Lambert. \textbf{Right.} Europa, the first telescope of the SPECULOOS Southern Observatory. Credit: P. Aniol.}
\label{fig:1}       
\end{figure}

We decided to initiate SPECULOOS first in the southern hemisphere, with a facility composed of four robotic Ritchey-Chretien (F/8) telescopes of 1-m diameter (see Fig.~\ref{fig:1}) currently being installed at ESO Paranal Observatory in the Chilean Atacama desert. The name of this facility is the SPECULOOS Southern Observatory\footnote{https://www.eso.org/public/teles-instr/paranal-observatory/speculoos/}. Each telescope is equipped with an Andor Peltier-cooled deeply depleted $\mathrm{2K\times2K}$ CCD camera with a 13.5~$\mu$m pixel size. The field of view of each telescope is $12~\arcm \times 12~\arcm$ and the corresponding pixel scale is $0.35\arcm\arcm$. This set-up is optimized to observe, in good seeing conditions (median seeing $<1.2\arcm\arcm$), with optimal sensitivity in the very-near-IR (700 to 1000~nm), UCDs with J-magnitude up to 14, and to obtain for them light curves with 0.1\% photometric precisions for a few minutes sampling times. In survey operations, each telescope will observe continuously one target during an average period of 10 nights (fine-tuned as a function of the spectral type). This duration of the observation sequence is optimized to explore efficiently the HZ of UCDs for transiting planets. To observe 500 targets in the southern hemisphere, a total of 12000 nights are needed, what can be done in 5 years considering realistic observations with four telescopes. All four telescopes are expected to be operation by the end of 2017.

The continuous observation of the targets does not only maximize the photon counts, it also improves the photometric detection threshold by letting the telescope keep the stars on the same pixels of the detector during the whole night. This continuous monitoring optimizes the capacity to detect low-amplitude transits, which is crucial here as all planetary formation models agree on the fact that UCDs should form small planets (e.g. \citealt{2009Icar..202....1M,2007ApJ...669..606R}). Furthermore, the need for continuous observation is driven by the expected short transit duration (down to 15~min) for planets orbiting at the inner edge of the habitable zone of UCDs \citep{2013ApJ...765..131K}. Once a transit signature is detected in the photometric data, the first follow-up action will be to confirm it by prolongating the SPECULOOS monitoring of the star, and possibly by using similar telescopes at other longitudes. Once a transit ephemeris will be secured, larger ground-based telescopes like the VLT, or space facilities like Spitzer (as was done in \citealt{2017Natur.542..456G}), may then be used to gather high precision transit photometry at different wavelengths, to assess the achromaticity of the eclipses and confirm their planetary origins \citep{2016Natur.533..221G}.

We are currently preparing a northern counterpart to the SPECULOOS Southern Observatory which should be operational in 2019. Its location is not decided yet. 

\section{SPECULOOS prototype on TRAPPIST}
Before the discovery of the TRAPPIST-1 planetary system, the relevance of a dedicated transit search targeting nearby UCDs could have been \textit{a priori}  questioned. The need for the individual monitoring of each UCD makes necessary  being able to detect a {\it single} transit event to prevent booking one telescope to one UCD for unrealistically long durations, and thus puts the strongest constraint on the required photometric precision. Related to this point, UCDs are faint and emit most of their light in the IR, \textit{a priori} suggesting the need for expensive large telescopes and IR detectors. SNR computations convinced us that it was not the case, and that telescopes of relatively modest sizes (60-cm to 2-m) equipped with near-IR-optimized CCD cameras should reach the required high photometric precisions. This had to be demonstrated. In addition to their intrinsic faintness, late M-dwarfs are commonly considered as active objects \citep{2005AN....326.1059G, 2005nlds.book.....R}. This activity could be a big issue, as it could strongly limit the ability to detect low amplitude transits (e.g., \citealt{2005nlds.book.....R}). Another possible barrier to the relevance of the SPECULOOS project concept could have come from the Earth's atmosphere itself. Indeed, in the very-near-IR, the water molecule and $\mathrm{OH}$ radical contribute to a number of absorption bands, as well as significant emission for $\mathrm{OH}$ (airglow). This brings unavoidable important levels of red noise in photometric time-series (e.g., \citealt{2011ApJ...736...12B}). For all these reasons, a thorough assessment study based on a prototype survey was mandatory.

\begin{figure}
\centering
\includegraphics[scale=.065]{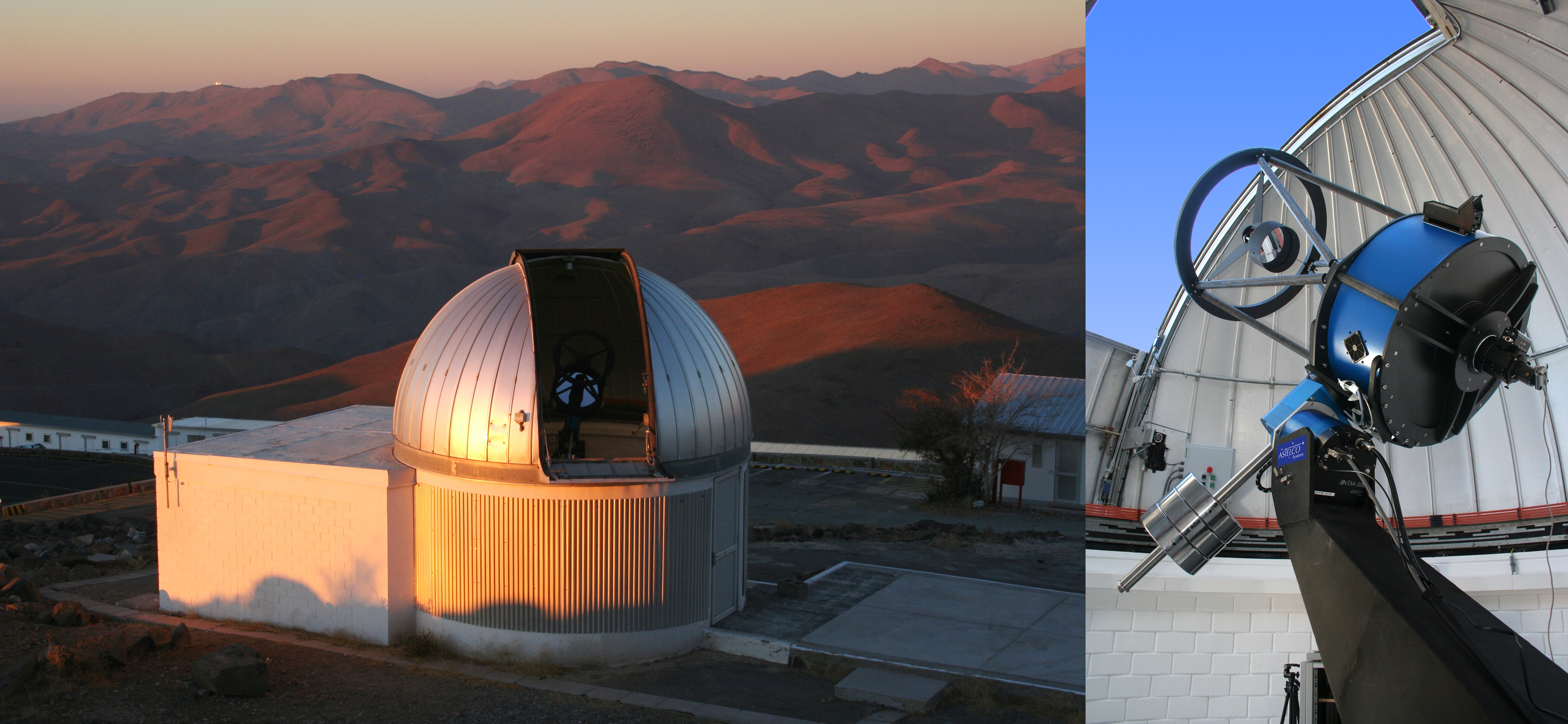}
\caption{\textbf{Left.} The dome of the TRAPPIST-South telescope at ESO La Silla Observatory (Chile). \textbf{Right.} TRAPPIST-South telescope. Credit: E.~Jehin.}
\label{fig:2}       
\end{figure}

In 2011, we initiated such a prototype survey for SPECULOOS with the robotic telescope TRAPPIST-South (\textbf{TRA}nsiting \textbf{P}lanets and \textbf{P}lanetes\textbf{I}mals \textbf{S}mall \textbf{T}elescope; \citealt{2011EPJWC..1106002G, 2011Msngr.145....2J}). It is a 60-cm (F/8) Ritchey-Chretien telescope installed by the University of Li\`ege in 2010 at ESO La Silla Observatory in the Atacama Desert in Chile (see Fig.~\ref{fig:2}). It is equipped with a near-IR optimized $\mathrm{2K\times2K}$ CCD camera with a $\mathrm{0.64~\arcm\arcm/pixel}$ scale, offering excellent quantum efficiencies from 300 to $>$~900~nm. This SPECULOOS prototype survey  targets 50 among the brightest southern UCDs, with J-magnitude between 5.4 and 12 (mean $\mathrm{J=11.3}$), and uniformly distributed in terms of spectral type and sky position. Its concept is to monitor in a wide near-IR filter (transmission $>90\%$ from 720~nm) each UCD during at least 100 hours spread over several nights. Its initial goals were to assess the typical photometric precisions that can be reached for UCDs on nightly timescales, the resulting detection thresholds for terrestrial planets, and to identify the astrophysical and atmospheric limitations of the SPECULOOS concept.

Since mid-2016 a northern extension of the prototype survey for SPECULOOS is conducted with the TRAPPIST-North telescope, installed at the Ouka\"imeden observatory in Morocco. This telescope is a twin-brother of the TRAPPIST-South telescope and operated in collaboration with the Cadi Ayyad University of Marrakesh.

Almost 40 UCDs were observed by TRAPPIST-South in the period from 2011 to 2016. Half of the observed UCDs show "flat" light curves, i.e. stable photometry on the night timescale. Some of the other UCDs ($\sim$20\%) show clear flares in some light curves. These flares are seen in near-IR light curves as sudden increase of a few percents of the measured brightness, followed by a gradual decrease back or close to the normal level. The whole process takes only 10 to 30~min. In the context of a transit search, it is easy to identify and discard the affected portions of light curves. Furthermore, their frequency is relatively small (1 flare per 3-4 nights on average). Finally, about 30\% of the observed UCDs show some rotational modulation (and more complex variability) with up to 5\% amplitude.

Within the framework of this prototype survey, we also monitored the nearby brown dwarf binary Luhman 16AB for nearly a fortnight, right after its discovery was announced in February 2013 \citep{2013ApJ...767L...1L}. The quality of our photometric data allowed us to reveal fast-evolving weather patterns in the atmosphere of the coolest component of the binary, as well as to firmly discard the transit of a two-Earth radius planet over the duration of the observations and of an Earth-sized planet on orbits shorter than $\sim$9.5 hours \citep{2013A&A...555L...5G}.

From intense simulations based on the injection and recovery of synthetic transits of terrestrial planets in actual TRAPPIST-South UCD light curves, we have reached the conclusion that the variability of a fraction of UCDs (flares and rotational modulation) does not limit the ability to detect transits of close-in planets. The reached photometric precisions are globally nominal. There is no hint of extra-amount of correlated noise, except for the small fraction ($\sim$10\%) of the observations performed in  high humidity conditions. Nominal sub-mmag precisions can thus be reached for UCDs from a suitable astronomical site (good transparency, low humidity). This conclusion was recently strengthened  by our detection of  Earth-sized exoplanets transiting one of the TRAPPIST-South UCD target, 2MASS J23062928-0502285 (TRAPPIST-1, \citealt{2016Natur.533..221G,2017Natur.542..456G}). The detection - and even the very existence - of this planetary system fully demonstrates the instrumental concept and the scientific potential of SPECULOOS.

\section{The TRAPPIST-1 planetary system}

\begin{figure}
\centering
\includegraphics[scale=.65]{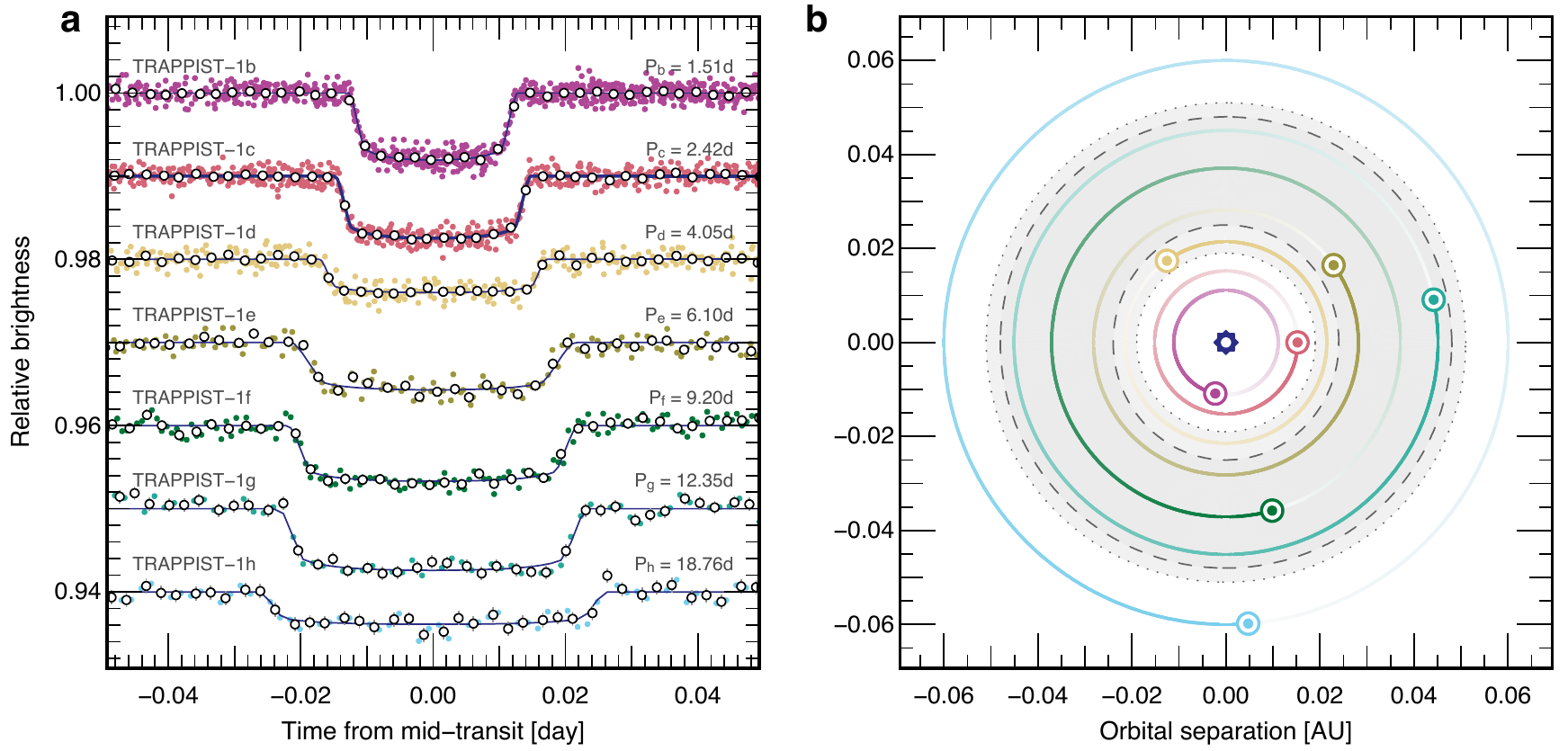}
\caption{\textbf{Panel a.} Period-folded transit light curves for the seven planets of TRAPPIST-1, resulting from the nearly-continuous observation of the star by the Spitzer space telescope from 19 September to 10 October 2016. The individual transits were corrected for the measured TTVs to produce this figure. Coloured dots show the unbinned measurements, whereas the open circles depict binned measurements for visual clarity. The 1-sigma error bars of the binned measurements are shown as vertical lines. The best-fit transit models are shown as coloured lines. \textbf{Panel b.} Representation of the orbits of the 7 planets. The same colour code as in panel a is used to identify the planets. The grey annulus and the two dashed lines represent the habitable zone of the star based on different theoretical assumptions (see \citealt{2017Natur.542..456G} and \citealt{2017arXiv170304166L} for details).}
\label{fig:3} 
\end{figure}

The TRAPPIST-1 planetary system is composed of at least seven planets with sizes and initial mass estimates similar to the Earth transiting an UCD star 39 light-years away \citep{2017Natur.542..456G}. The host star is a moderately active $\mathrm{M8\pm0.5}$ dwarf star (V=18.80, R=16.47, I=14.02, J=11.35, K=10.30). Its mass, radius and temperature are estimated to be  $\mathrm{0.0802~\pm~0.0073~M_{\odot}}$, $\mathrm{0.117\pm0.0036~R_{\odot}}$, and $\mathrm{2559\pm50~K}$, respectively. Fig.~\ref{fig:3}a shows the transit light curves of the planets as observed at 4.5$\mu$ by Spitzer, while a representation of their orbits is shown in Fig.~\ref{fig:3}b. These seven planets form a unique near-resonant chain such that their orbital periods (1.51, 2.42, 4.04, 6.06, 9.21, 12.35 and 18.76 days) are near-ratios of small integers \citep{2017Natur.542..456G, 2017arXiv170304166L}. Transit Time Variation (TTV) signals from a few tens of seconds to more than 30 minutes indicate significant mutual interactions between the planets \citep{2005MNRAS.359..567A,2005Sci...307.1288H,2010exop.book..217F}. By analysing TTV signals, we could determine initial mass estimates for the six inner planets, along with upper limits on their orbital eccentricities (e $< 0.085$). 

As shown in Fig.~\ref{fig:4}a, the stellar irradiations of the planets cover a range from $\sim$4.3 to \hbox{$\sim$0.14 $\mathrm{S_{E}}$} \hbox{($\mathrm{S_{E}}$ = Solar} irradiation at 1~au), which is very similar to the range seen in the inner Solar System (Mercury = 6.7~$\mathrm{S_{E}}$ , Ceres = 0.13~$\mathrm{S_{E}}$). The equilibrium temperature of the outermost detected planet, TRAPPIST-1h, is 170~K (assuming a null albedo), placing it at the snow line of the system \citep{2017arXiv170304166L}.  The derived planets' orbital inclinations are all very close to 90$^{\circ}$, indicating a dramatically coplanar system seen nearly edge-on. This architecture, combined with the fact that the planets form a near-resonant chain, suggests that they formed farther from the star and migrated inward through interactions with the disc (e.g. \citealt{2006A&A...450..833C, 2005MNRAS.363..153P, 2007ApJ...654.1110T}). Planets TRAPPIST-1e, f, and g are firmly in the HZ of the star (as estimated following \citealt{2013ApJ...765..131K}, see Fig.~\ref{fig:3}b). They are thus prime targets for the search of potential atmospheric biosignatures with the next generation of telescopes and instruments.


\section{Characterization of TRAPPIST-1 planets: present and future}

The transiting configuration of the TRAPPIST-1 planets, combined with the small size (0.12~$\mathrm{R_{\odot}}$), low luminosity (0.0005~$\mathrm{L_{\odot}}$), and infrared brightness ($\mathrm{K=10.3}$) of their host star, provides the extraordinary opportunity to thoroughly study their atmospheric properties with present-day and future astronomical facilities. Our \textit{a priori} knowledge about their atmospheric compositions is very limited as they are the first transiting planets detected around a UCD. Theoretical predictions span the entire atmospheric range, from extended H/He-dominated to depleted atmospheres \citep{2013ApJ...775..105O, 2014ApJ...795...65J, 2015ApJ...815L..12J, 2015AsBio..15..119L, 2015ExA....40..449L, 2016MNRAS.459.4088O}. A cloud-free H/He-dominated atmosphere should yield prominent spectroscopic signatures of H$_2$O and/or CH$_4$ in the near-infrared, readily detectable with current space-based instrumentation through transit transmission spectroscopy. Reconnaissance observations of the innermost planets b and c during transit conjunction with HST/WFC3 ruled out this scenario for these two planets \citep{2016Natur.537...69D}, but still allow for many cloudy and/or denser atmospheres, such as H$_2$O-, N$_2$-, or CO$_2$-dominated atmospheres. Similar observations have been obtained to assess the atmospheres of planets d, e, f, and g as well (de Wit et al. submitted). Further spectroscopic transmission observations of the TRAPPIST-1 planets in the infrared will allow for progressively higher mean molecular weight atmospheres to be probed, before their in-depth characterization with JWST \citep{2016MNRAS.461L..92B} and the next-generation of ground-based extremely large telescopes (E-ELT, GMT, and TMT).

Efforts are also ongoing to characterize the high-energy radiation environment of the planets, which is a key factor in assessing their atmospheric stability and potential habitability. XMM-Newton observations showed that TRAPPIST-1 is a relatively strong and variable coronal X-ray source with an X-ray luminosity similar to that of the quiet Sun, despite a much lower bolometric luminosity \citep{2017MNRAS.465L..74W}. The Ly-$\alpha$ line of TRAPPIST-1 was also recently measured using HST/STIS and was found to be much fainter than expected from the X-ray emission, which may suggest that TRAPPIST-1 chromosphere is moderately active compared to its transition region and corona \citep{2017A&A...599L...3B}. Still, TRAPPIST-1 \hbox{Ly-$\alpha$} line is bright enough to perform transit spectroscopy and search for signatures of planetary hydrogen escape. Hydrogen exospheres could indicate the presence of water-vapor being photo-dissociated in the upper atmospheres and replenished by evaporating water oceans, thus hinting at large water reservoirs on the planets \citep{2004ApJ...605L..65J}. Theoretical models predict that the total XUV irradiation derived from the XMM-Newton and HST/STIS data could be strong enough to strip Earth-like atmospheres and oceans from planets b and c in approximately 1 and 3 Gyr, while the same process would take between 5 and 22 Gyr for planets d to g \citep{2017MNRAS.464.3728B, 2017A&A...599L...3B}. However, the observed high-energy fluxes are highly variable and these first observations may not provide a complete picture of the high-energy radiation environment of the TRAPPIST-1 planets, making it necessary to gather additional measurements.

\begin{figure}
\centering
\includegraphics[scale=.45]{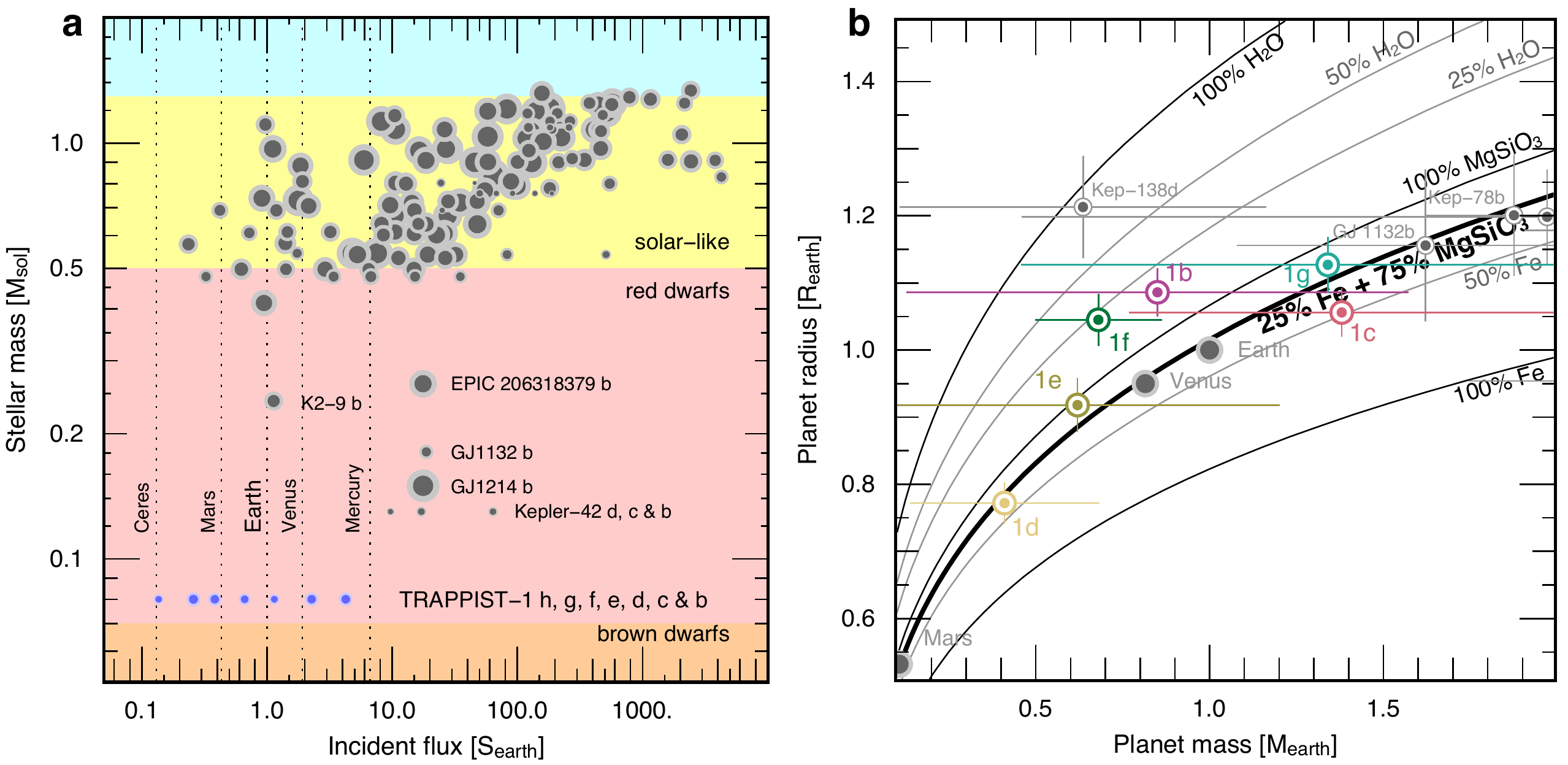}

\caption{\textbf{Panel a.} Masses of host stars and incident fluxes of known sub-Neptune-sized exoplanets. The size of the symbols scales linearly with the radius of the planet. The background is colour-coded according
to stellar mass (in units of the Sun's mass). The TRAPPIST-1 planets, shown in blue, are at the boundary between planets associated with hydrogen-burning stars and planets associated with brown dwarfs. The positions of the Solar System terrestrial planets and Ceres are shown for reference. Only the exoplanets with a measured radius equal to or smaller than that of GJ 1214b are included. \textbf{Panel b.} Mass-radius diagram for terrestrial planets. The coloured circular symbols with error bars represent the TRAPPIST-1 planets. The solid lines are theoretical mass-radius curves for planets with different compositions from \cite{2016ApJ...819..127Z}. Figure from \cite{2017Natur.542..456G}.}

\label{fig:4} 
\end{figure}

Precise mass determinations and the resulting constraints on their compositions (e.g. \citealt{2016ApJ...819..127Z}) are critical for a thorough understanding of these planets, as well as for the optimal exploitation of future atmospheric observations \citep{2013Sci...342.1473D}. Assessment of the planets' potential habitability makes it also necessary to measure their orbital eccentricities, to constrain the impact of tidal heating on their total energy budget, but also on their geological activity that could be able to counterbalance atmospheric erosion through volcanism. The current mass estimates for the six inner planets broadly suggest rocky compositions (see Fig.~\ref{fig:4}b), but are at present too uncertain to constrain the fraction of volatiles in the planets' compositions, except for planet f, whose low density suggests a volatile-rich composition. We also find small but non-zero eccentricities, resulting in significant tidal heating: all planets except f and h have a tidal heat flux higher than Earth's total heat flux \citep{2017arXiv170304166L}. However, this global dynamical solution is preliminary and may not correspond to a global minimum of the parameter space, as several solutions fit equally well our currently limited dataset. Additional precise transit timings for all seven planets will be key in constraining further the planet masses and eccentricities and in isolating a unique, well defined, dynamical solution.

The TRAPPIST-1 system is a rich laboratory for exploring the factors relevant to exoplanet habitability and a blueprint for investigations of the many planetary systems expected to be found by SPECULOOS. Observations of these systems with JWST and other upcoming facilities will open the era of detailed characterization of temperate terrestrial planets, and will provide us with first opportunities to detect chemical traces of life beyond our Solar System.

\begin{acknowledgement}
The authors are grateful and thank Adam Burgasser, Amaury H. M. J. Triaud, Catarina S. Fernandes, Julien de Wit, and the rest of SPECULOOS and TRAPPIST teams for the help in the preparation of this chapter. TRAPPIST-South is a project funded by the Belgian Fonds (National) de la Recherche Scientifique (F.R.S.-FNRS) under grant FRFC 2.5.594.09.F, with the participation of the Swiss National Science Foundation (FNS/SNSF). TRAPPIST-North is a project funded by the University of Li\`ege, and performed in collaboration with Cadi Ayyad University of Marrakesh. The research leading to these results has received funding from the European Research Council (ERC) under the FP/2007-2013 ERC grant agreement no. 336480, and under the H2020 ERC grant agreement no. 679030; and from an Actions de Recherche Concert\'ee (ARC) grant, financed by the Wallonia-Brussels Federation. This work was also partially supported by a grant from the Simons Foundation (PI Queloz, grant number 327127). L.D. acknowledges support from the Gruber Foundation Fellowship. M.G. and E.J. are F.R.S.-FNRS research associates.
\end{acknowledgement}

\bibliographystyle{spbasicHBexo}  
\bibliography{burdanov} 

\begin{thebibliography}{81}
\providecommand{\natexlab}[1]{#1}
\providecommand{\url}[1]{{#1}}
\providecommand{\urlprefix}{URL }
\expandafter\ifx\csname urlstyle\endcsname\relax
  \providecommand{\doi}[1]{DOI~\discretionary{}{}{}#1}\else
  \providecommand{\doi}{DOI~\discretionary{}{}{}\begingroup
  \urlstyle{rm}\Url}\fi
\providecommand{\eprint}[2][]{\url{#2}}

\bibitem[{{Agol} et~al.(2005){Agol}, {Steffen}, {Sari}, and
  {Clarkson}}]{2005MNRAS.359..567A}
{Agol} E, {Steffen} J, {Sari} R {Clarkson} W (2005) {On detecting terrestrial
  planets with timing of giant planet transits}. \mnras 359:567--579

\bibitem[{{Alibert} and {Benz}(2017)}]{2017A&A...598L...5A}
{Alibert} Y {Benz} W (2017) {Formation and composition of planets around very
  low mass stars}. \aap 598:L5

\bibitem[{{Bakos} et~al.(2007){Bakos}, {Noyes}, {Kov{\'a}cs}, {Latham},
  {Sasselov}, {Torres}, {Fischer}, {Stefanik}, {Sato}, {Johnson}, {P{\'a}l},
  {Marcy}, {Butler}, {Esquerdo}, {Stanek}, {L{\'a}z{\'a}r}, {Papp}, {S{\'a}ri},
  and {Sip{\H o}cz}}]{2007ApJ...656..552B}
{Bakos} G{\'A}, {Noyes} RW, {Kov{\'a}cs} G et~al. (2007) {HAT-P-1b: A
  Large-Radius, Low-Density Exoplanet Transiting One Member of a Stellar
  Binary}. \apj 656:552--559

\bibitem[{{Ballard} and {Johnson}(2016)}]{2016ApJ...816...66B}
{Ballard} S {Johnson} JA (2016) {The Kepler Dichotomy among the M Dwarfs: Half
  of Systems Contain Five or More Coplanar Planets}. \apj 816:66

\bibitem[{{Barstow} and {Irwin}(2016)}]{2016MNRAS.461L..92B}
{Barstow} JK {Irwin} PGJ (2016) {Habitable worlds with JWST: transit
  spectroscopy of the TRAPPIST-1 system?} \mnras 461:L92--L96

\bibitem[{{Belu} et~al.(2011){Belu}, {Selsis}, {Morales}, {Ribas}, {Cossou},
  and {Rauer}}]{2011A&A...525A..83B}
{Belu} AR, {Selsis} F, {Morales} JC et~al. (2011) {Primary and secondary
  eclipse spectroscopy with JWST: exploring the exoplanet parameter space}.
  \aap 525:A83

\bibitem[{{Berger}(2006)}]{2006ApJ...648..629B}
{Berger} E (2006) {Radio Observations of a Large Sample of Late M, L, and T
  Dwarfs: The Distribution of Magnetic Field Strengths}. \apj 648:629--636

\bibitem[{{Berta} et~al.(2011){Berta}, {Charbonneau}, {Bean}, {Irwin}, {Burke},
  {D{\'e}sert}, {Nutzman}, and {Falco}}]{2011ApJ...736...12B}
{Berta} ZK, {Charbonneau} D, {Bean} J et~al. (2011) {The GJ1214 Super-Earth
  System: Stellar Variability, New Transits, and a Search for Additional
  Planets}. \apj 736:12

\bibitem[{{Blake} et~al.(2010){Blake}, {Charbonneau}, and
  {White}}]{2010ApJ...723..684B}
{Blake} CH, {Charbonneau} D {White} RJ (2010) {The NIRSPEC Ultracool Dwarf
  Radial Velocity Survey}. \apj 723:684--706

\bibitem[{{Bolmont} et~al.(2011){Bolmont}, {Raymond}, and
  {Leconte}}]{2011A&A...535A..94B}
{Bolmont} E, {Raymond} SN {Leconte} J (2011) {Tidal evolution of planets around
  brown dwarfs}. \aap 535:A94

\bibitem[{{Bolmont} et~al.(2017){Bolmont}, {Selsis}, {Owen}, {Ribas},
  {Raymond}, {Leconte}, and {Gillon}}]{2017MNRAS.464.3728B}
{Bolmont} E, {Selsis} F, {Owen} JE et~al. (2017) {Water loss from terrestrial
  planets orbiting ultracool dwarfs: implications for the planets of
  TRAPPIST-1}. \mnras 464:3728--3741

\bibitem[{{Bourrier} et~al.(2017){Bourrier}, {Ehrenreich}, {Wheatley},
  {Bolmont}, {Gillon}, {de Wit}, {Burgasser}, {Jehin}, {Queloz}, and
  {Triaud}}]{2017A&A...599L...3B}
{Bourrier} V, {Ehrenreich} D, {Wheatley} PJ et~al. (2017) {Reconnaissance of
  the TRAPPIST-1 exoplanet system in the Lyman-{$\alpha$} line}. \aap 599:L3

\bibitem[{{Burgasser} et~al.(2015){Burgasser}, {Gillon}, {Melis}, {Bowler},
  {Michelsen}, {Bardalez Gagliuffi}, {Gelino}, {Jehin}, {Delrez}, {Manfroid},
  and {Blake}}]{2015AJ....149..104B}
{Burgasser} AJ, {Gillon} M, {Melis} C et~al. (2015) {WISE J072003.20-084651.2:
  an Old and Active M9.5 + T5 Spectral Binary 6 pc from the Sun}. \aj 149:104

\bibitem[{{Burrows} and {Liebert}(1993)}]{1993RvMP...65..301B}
{Burrows} A {Liebert} J (1993) {The science of brown dwarfs}. Reviews of Modern
  Physics 65:301--336

\bibitem[{{Collier Cameron} et~al.(2007){Collier Cameron}, {Bouchy},
  {H{\'e}brard}, {Maxted}, {Pollacco}, {Pont}, {Skillen}, {Smalley}, {Street},
  {West}, {Wilson}, {Aigrain}, {Christian}, {Clarkson}, {Enoch}, {Evans},
  {Fitzsimmons}, {Fleenor}, {Gillon}, {Haswell}, {Hebb}, {Hellier}, {Hodgkin},
  {Horne}, {Irwin}, {Kane}, {Keenan}, {Loeillet}, {Lister}, {Mayor}, {Moutou},
  {Norton}, {Osborne}, {Parley}, {Queloz}, {Ryans}, {Triaud}, {Udry}, and
  {Wheatley}}]{2007MNRAS.375..951C}
{Collier Cameron} A, {Bouchy} F, {H{\'e}brard} G et~al. (2007) {WASP-1b and
  WASP-2b: two new transiting exoplanets detected with SuperWASP and SOPHIE}.
  \mnras 375:951--957

\bibitem[{{Cresswell} and {Nelson}(2006)}]{2006A&A...450..833C}
{Cresswell} P {Nelson} RP (2006) {On the evolution of multiple protoplanets
  embedded in a protostellar disc}. \aap 450:833--853

\bibitem[{{Cushing} et~al.(2011){Cushing}, {Kirkpatrick}, {Gelino}, {Griffith},
  {Skrutskie}, {Mainzer}, {Marsh}, {Beichman}, {Burgasser}, {Prato}, {Simcoe},
  {Marley}, {Saumon}, {Freedman}, {Eisenhardt}, and
  {Wright}}]{2011ApJ...743...50C}
{Cushing} MC, {Kirkpatrick} JD, {Gelino} CR et~al. (2011) {The Discovery of Y
  Dwarfs using Data from the Wide-field Infrared Survey Explorer (WISE)}. \apj
  743:50

\bibitem[{{de Wit} and {Seager}(2013)}]{2013Sci...342.1473D}
{de Wit} J {Seager} S (2013) {Constraining Exoplanet Mass from Transmission
  Spectroscopy}. Science 342:1473--1477

\bibitem[{{de Wit} et~al.(2016){de Wit}, {Wakeford}, {Gillon}, {Lewis},
  {Valenti}, {Demory}, {Burgasser}, {Burdanov}, {Delrez}, {Jehin}, {Lederer},
  {Queloz}, {Triaud}, and {Van Grootel}}]{2016Natur.537...69D}
{de Wit} J, {Wakeford} HR, {Gillon} M et~al. (2016) {A combined transmission
  spectrum of the Earth-sized exoplanets TRAPPIST-1 b and c}. \nat 537:69--72

\bibitem[{{Deming} and {Seager}(2017)}]{2017arXiv170100493D}
{Deming} D {Seager} S (2017) {Illusion and Reality in the Atmospheres of
  Exoplanets}. ArXiv e-prints

\bibitem[{{Dieterich} et~al.(2014){Dieterich}, {Henry}, {Jao}, {Winters},
  {Hosey}, {Riedel}, and {Subasavage}}]{2014AJ....147...94D}
{Dieterich} SB, {Henry} TJ, {Jao} WC et~al. (2014) {The Solar Neighborhood.
  XXXII. The Hydrogen Burning Limit}. \aj 147:94

\bibitem[{{Fabrycky}(2010)}]{2010exop.book..217F}
{Fabrycky} DC (2010) {Non-Keplerian Dynamics of Exoplanets}, pp 217--238

\bibitem[{{Figueira} et~al.(2012){Figueira}, {Marmier}, {Bou{\'e}}, {Lovis},
  {Santos}, {Montalto}, {Udry}, {Pepe}, and {Mayor}}]{2012A&A...541A.139F}
{Figueira} P, {Marmier} M, {Bou{\'e}} G et~al. (2012) {Comparing HARPS and
  Kepler surveys. The alignment of multiple-planet systems}. \aap 541:A139

\bibitem[{{Gillon} et~al.(2011){Gillon}, {Jehin}, {Magain}, {Chantry},
  {Hutsem{\'e}kers}, {Manfroid}, {Queloz}, and {Udry}}]{2011EPJWC..1106002G}
{Gillon} M, {Jehin} E, {Magain} P et~al. (2011) {TRAPPIST: a robotic telescope
  dedicated to the study of planetary systems}. In: European Physical Journal
  Web of Conferences, European Physical Journal Web of Conferences, vol~11, p
  06002, \doi{10.1051/epjconf/20101106002}

\bibitem[{{Gillon} et~al.(2013){Gillon}, {Triaud}, {Jehin}, {Delrez}, {Opitom},
  {Magain}, {Lendl}, and {Queloz}}]{2013A&A...555L...5G}
{Gillon} M, {Triaud} AHMJ, {Jehin} E et~al. (2013) {Fast-evolving weather for
  the coolest of our two new substellar neighbours}. \aap 555:L5

\bibitem[{{Gillon} et~al.(2016){Gillon}, {Jehin}, {Lederer}, {Delrez}, {de
  Wit}, {Burdanov}, {Van Grootel}, {Burgasser}, {Triaud}, {Opitom}, {Demory},
  {Sahu}, {Bardalez Gagliuffi}, {Magain}, and {Queloz}}]{2016Natur.533..221G}
{Gillon} M, {Jehin} E, {Lederer} SM et~al. (2016) {Temperate Earth-sized
  planets transiting a nearby ultracool dwarf star}. \nat 533:221--224

\bibitem[{{Gillon} et~al.(2017){Gillon}, {Triaud}, {Demory}, {Jehin}, {Agol},
  {Deck}, {Lederer}, {de Wit}, {Burdanov}, {Ingalls}, {Bolmont}, {Leconte},
  {Raymond}, {Selsis}, {Turbet}, {Barkaoui}, {Burgasser}, {Burleigh}, {Carey},
  {Chaushev}, {Copperwheat}, {Delrez}, {Fernandes}, {Holdsworth}, {Kotze}, {Van
  Grootel}, {Almleaky}, {Benkhaldoun}, {Magain}, and
  {Queloz}}]{2017Natur.542..456G}
{Gillon} M, {Triaud} AHMJ, {Demory} BO et~al. (2017) {Seven temperate
  terrestrial planets around the nearby ultracool dwarf star TRAPPIST-1}. \nat
  542:456--460

\bibitem[{{Gizis} et~al.(2000){Gizis}, {Monet}, {Reid}, {Kirkpatrick},
  {Liebert}, and {Williams}}]{2000AJ....120.1085G}
{Gizis} JE, {Monet} DG, {Reid} IN et~al. (2000) {New Neighbors from 2MASS:
  Activity and Kinematics at the Bottom of the Main Sequence}. \aj
  120:1085--1099

\bibitem[{{Goldman}(2005)}]{2005AN....326.1059G}
{Goldman} B (2005) {Ultra-cool dwarf variability}. Astronomische Nachrichten
  326:1059--1064

\bibitem[{{Han} et~al.(2014){Han}, {Wang}, {Wright}, {Feng}, {Zhao},
  {Fakhouri}, {Brown}, and {Hancock}}]{2014PASP..126..827H}
{Han} E, {Wang} SX, {Wright} JT et~al. (2014) {Exoplanet Orbit Database. II.
  Updates to Exoplanets.org}. \pasp 126:827

\bibitem[{{Hayashi} and {Nakano}(1963)}]{1963PThPh..30..460H}
{Hayashi} C {Nakano} T (1963) {Evolution of Stars of Small Masses in the
  Pre-Main-Sequence Stages}. Progress of Theoretical Physics 30:460--474

\bibitem[{{He} et~al.(2017){He}, {Triaud}, and {Gillon}}]{2017MNRAS.464.2687H}
{He} MY, {Triaud} AHMJ {Gillon} M (2017) {First limits on the occurrence rate
  of short-period planets orbiting brown dwarfs}. \mnras 464:2687--2697

\bibitem[{{Holman} and {Murray}(2005)}]{2005Sci...307.1288H}
{Holman} MJ {Murray} NW (2005) {The Use of Transit Timing to Detect
  Terrestrial-Mass Extrasolar Planets}. Science 307:1288--1291

\bibitem[{{Jehin} et~al.(2011){Jehin}, {Gillon}, {Queloz}, {Magain},
  {Manfroid}, {Chantry}, {Lendl}, {Hutsem{\'e}kers}, and
  {Udry}}]{2011Msngr.145....2J}
{Jehin} E, {Gillon} M, {Queloz} D et~al. (2011) {TRAPPIST: TRAnsiting Planets
  and PlanetesImals Small Telescope}. The Messenger 145:2--6

\bibitem[{{Jin} et~al.(2014){Jin}, {Mordasini}, {Parmentier}, {van Boekel},
  {Henning}, and {Ji}}]{2014ApJ...795...65J}
{Jin} S, {Mordasini} C, {Parmentier} V et~al. (2014) {Planetary Population
  Synthesis Coupled with Atmospheric Escape: A Statistical View of
  Evaporation}. \apj 795:65

\bibitem[{{Johnstone} et~al.(2015){Johnstone}, {G{\"u}del}, {St{\"o}kl},
  {Lammer}, {Tu}, {Kislyakova}, {L{\"u}ftinger}, {Odert}, {Erkaev}, and
  {Dorfi}}]{2015ApJ...815L..12J}
{Johnstone} CP, {G{\"u}del} M, {St{\"o}kl} A et~al. (2015) {The Evolution of
  Stellar Rotation and the Hydrogen Atmospheres of Habitable-zone Terrestrial
  Planets}. \apjl 815:L12

\bibitem[{{Jura}(2004)}]{2004ApJ...605L..65J}
{Jura} M (2004) {An Observational Signature of Evolved Oceans on Extrasolar
  Terrestrial Planets}. \apjl 605:L65--L68

\bibitem[{{Kaltenegger} and {Traub}(2009)}]{2009ApJ...698..519K}
{Kaltenegger} L {Traub} WA (2009) {Transits of Earth-like Planets}. \apj
  698:519--527

\bibitem[{{Kirkpatrick}(2005)}]{2005ARA&A..43..195K}
{Kirkpatrick} JD (2005) {New Spectral Types L and T}. \araa 43:195--245

\bibitem[{{Kirkpatrick} et~al.(2012){Kirkpatrick}, {Gelino}, {Cushing}, {Mace},
  {Griffith}, {Skrutskie}, {Marsh}, {Wright}, {Eisenhardt}, {McLean},
  {Mainzer}, {Burgasser}, {Tinney}, {Parker}, and
  {Salter}}]{2012ApJ...753..156K}
{Kirkpatrick} JD, {Gelino} CR, {Cushing} MC et~al. (2012) {Further Defining
  Spectral Type ``Y'' and Exploring the Low-mass End of the Field Brown Dwarf
  Mass Function}. \apj 753:156

\bibitem[{{Klein} et~al.(2003){Klein}, {Apai}, {Pascucci}, {Henning}, and
  {Waters}}]{2003ApJ...593L..57K}
{Klein} R, {Apai} D, {Pascucci} I, {Henning} T {Waters} LBFM (2003) {First
  Detection of Millimeter Dust Emission from Brown Dwarf Disks}. \apjl
  593:L57--L60

\bibitem[{{Kopparapu} et~al.(2013){Kopparapu}, {Ramirez}, {Kasting}, {Eymet},
  {Robinson}, {Mahadevan}, {Terrien}, {Domagal-Goldman}, {Meadows}, and
  {Deshpande}}]{2013ApJ...765..131K}
{Kopparapu} RK, {Ramirez} R, {Kasting} JF et~al. (2013) {Habitable Zones around
  Main-sequence Stars: New Estimates}. \apj 765:131

\bibitem[{{Kubas} et~al.(2012){Kubas}, {Beaulieu}, {Bennett}, {Cassan}, {Cole},
  {Lunine}, {Marquette}, {Dong}, {Gould}, {Sumi}, {Batista}, {Fouqu{\'e}},
  {Brillant}, {Dieters}, {Coutures}, {Greenhill}, {Bond}, {Nagayama},
  {Udalski}, {Pompei}, {N{\"u}rnberger}, and {Le
  Bouquin}}]{2012A&A...540A..78K}
{Kubas} D, {Beaulieu} JP, {Bennett} DP et~al. (2012) {A frozen super-Earth
  orbiting a star at the bottom of the main sequence}. \aap 540:A78

\bibitem[{{Kumar}(1962)}]{1962AJ.....67S.579K}
{Kumar} SS (1962) {Study of Degeneracy in Very Light Stars.} \aj 67:579

\bibitem[{{Kumar}(1963)}]{1963ApJ...137.1121K}
{Kumar} SS (1963) {The Structure of Stars of Very Low Mass.} \apj 137:1121

\bibitem[{{Leconte} et~al.(2015){Leconte}, {Forget}, and
  {Lammer}}]{2015ExA....40..449L}
{Leconte} J, {Forget} F {Lammer} H (2015) {On the (anticipated) diversity of
  terrestrial planet atmospheres}. Experimental Astronomy 40:449--467

\bibitem[{{Luger} and {Barnes}(2015)}]{2015AsBio..15..119L}
{Luger} R {Barnes} R (2015) {Extreme Water Loss and Abiotic O2Buildup on
  Planets Throughout the Habitable Zones of M Dwarfs}. Astrobiology 15:119--143

\bibitem[{{Luger} et~al.(2017){Luger}, {Sestovic}, {Kruse}, {Grimm}, {Demory},
  {Agol}, {Bolmont}, {Fabrycky}, {Fernandes}, {Van Grootel}, {Burgasser},
  {Gillon}, {Ingalls}, {Jehin}, {Raymond}, {Selsis}, {Triaud}, {Barclay},
  {Barentsen}, {Delrez}, {de Wit}, {Foreman-Mackey}, {Holdsworth}, {Leconte},
  {Lederer}, {Turbet}, {Almleaky}, {Benkhaldoun}, {Magain}, {Morris}, {Heng},
  and {Queloz}}]{2017arXiv170304166L}
{Luger} R, {Sestovic} M, {Kruse} E et~al. (2017) {A terrestrial-sized exoplanet
  at the snow line of TRAPPIST-1}. ArXiv e-prints

\bibitem[{{Luhman}(2012)}]{2012ARA&A..50...65L}
{Luhman} KL (2012) {The Formation and Early Evolution of Low-Mass Stars and
  Brown Dwarfs}. \araa 50:65--106

\bibitem[{{Luhman}(2013)}]{2013ApJ...767L...1L}
{Luhman} KL (2013) {Discovery of a Binary Brown Dwarf at 2 pc from the Sun}.
  \apjl 767:L1

\bibitem[{{Luhman}(2014)}]{2014ApJ...786L..18L}
{Luhman} KL (2014) {Discovery of a \~{}250 K Brown Dwarf at 2 pc from the Sun}.
  \apjl 786:L18

\bibitem[{{Luhman} et~al.(2007){Luhman}, {Joergens}, {Lada}, {Muzerolle},
  {Pascucci}, and {White}}]{2007prpl.conf..443L}
{Luhman} KL, {Joergens} V, {Lada} C et~al. (2007) {The Formation of Brown
  Dwarfs: Observations}. Protostars and Planets V pp 443--457

\bibitem[{{Mamajek} et~al.(2015){Mamajek}, {Barenfeld}, {Ivanov}, {Kniazev},
  {V{\"a}is{\"a}nen}, {Beletsky}, and {Boffin}}]{2015ApJ...800L..17M}
{Mamajek} EE, {Barenfeld} SA, {Ivanov} VD et~al. (2015) {The Closest Known
  Flyby of a Star to the Solar System}. \apjl 800:L17

\bibitem[{{Mayor} and {Queloz}(1995)}]{1995Natur.378..355M}
{Mayor} M {Queloz} D (1995) {A Jupiter-mass companion to a solar-type star}.
  \nat 378:355--359

\bibitem[{{Metchev} et~al.(2015){Metchev}, {Heinze}, {Apai}, {Flateau},
  {Radigan}, {Burgasser}, {Marley}, {Artigau}, {Plavchan}, and
  {Goldman}}]{2015ApJ...799..154M}
{Metchev} SA, {Heinze} A, {Apai} D et~al. (2015) {Weather on Other Worlds. II.
  Survey Results: Spots are Ubiquitous on L and T Dwarfs}. \apj 799:154

\bibitem[{{Montgomery} and {Laughlin}(2009)}]{2009Icar..202....1M}
{Montgomery} R {Laughlin} G (2009) {Formation and detection of Earth mass
  planets around low mass stars}. \icarus 202:1--11

\bibitem[{{Muirhead} et~al.(2012){Muirhead}, {Johnson}, {Apps}, {Carter},
  {Morton}, {Fabrycky}, {Pineda}, {Bottom}, {Rojas-Ayala}, {Schlawin},
  {Hamren}, {Covey}, {Crepp}, {Stassun}, {Pepper}, {Hebb}, {Kirby}, {Howard},
  {Isaacson}, {Marcy}, {Levitan}, {Diaz-Santos}, {Armus}, and
  {Lloyd}}]{2012ApJ...747..144M}
{Muirhead} PS, {Johnson} JA, {Apps} K et~al. (2012) {Characterizing the Cool
  KOIs. III. KOI 961: A Small Star with Large Proper Motion and Three Small
  Planets}. \apj 747:144

\bibitem[{{Muirhead} et~al.(2015){Muirhead}, {Mann}, {Vanderburg}, {Morton},
  {Kraus}, {Ireland}, {Swift}, {Feiden}, {Gaidos}, and
  {Gazak}}]{2015ApJ...801...18M}
{Muirhead} PS, {Mann} AW, {Vanderburg} A et~al. (2015) {Kepler-445, Kepler-446
  and the Occurrence of Compact Multiples Orbiting Mid-M Dwarf Stars}. \apj
  801:18

\bibitem[{{Muzerolle} et~al.(2000){Muzerolle}, {Brice{\~n}o}, {Calvet},
  {Hartmann}, {Hillenbrand}, and {Gullbring}}]{2000ApJ...545L.141M}
{Muzerolle} J, {Brice{\~n}o} C, {Calvet} N et~al. (2000) {Detection of Disk
  Accretion at the Substellar Limit}. \apjl 545:L141--L144

\bibitem[{{Owen} and {Mohanty}(2016)}]{2016MNRAS.459.4088O}
{Owen} JE {Mohanty} S (2016) {Habitability of terrestrial-mass planets in the
  HZ of M Dwarfs - I. H/He-dominated atmospheres}. \mnras 459:4088--4108

\bibitem[{{Owen} and {Wu}(2013)}]{2013ApJ...775..105O}
{Owen} JE {Wu} Y (2013) {Kepler Planets: A Tale of Evaporation}. \apj 775:105

\bibitem[{{Papaloizou} and {Szuszkiewicz}(2005)}]{2005MNRAS.363..153P}
{Papaloizou} JCB {Szuszkiewicz} E (2005) {On the migration-induced resonances
  in a system of two planets with masses in the Earth mass range}. \mnras
  363:153--176

\bibitem[{{Pascucci} et~al.(2011){Pascucci}, {Laughlin}, {Gaudi}, {Kennedy},
  {Luhman}, {Mohanty}, {Birkby}, {Ercolano}, {Plavchan}, and
  {Skemer}}]{2011ASPC..448..469P}
{Pascucci} I, {Laughlin} G, {Gaudi} BS et~al. (2011) {Planet Formation Around
  M-dwarf Stars: From Young Disks to Planets}. In: {Johns-Krull} C, {Browning}
  MK {West} AA (eds) 16th Cambridge Workshop on Cool Stars, Stellar Systems,
  and the Sun, Astronomical Society of the Pacific Conference Series, vol 448,
  p 469

\bibitem[{{Payne} and {Lodato}(2007)}]{2007MNRAS.381.1597P}
{Payne} MJ {Lodato} G (2007) {The potential for Earth-mass planet formation
  around brown dwarfs}. \mnras 381:1597--1606

\bibitem[{{Pineda} et~al.(2016){Pineda}, {Hallinan}, {Kirkpatrick}, {Cotter},
  {Kao}, and {Mooley}}]{2016ApJ...826...73P}
{Pineda} JS, {Hallinan} G, {Kirkpatrick} JD et~al. (2016) {A Survey for
  H{$\alpha$} Emission from Late L Dwarfs and T Dwarfs}. \apj 826:73

\bibitem[{{Raymond} et~al.(2007){Raymond}, {Scalo}, and
  {Meadows}}]{2007ApJ...669..606R}
{Raymond} SN, {Scalo} J {Meadows} VS (2007) {A Decreased Probability of
  Habitable Planet Formation around Low-Mass Stars}. \apj 669:606--614

\bibitem[{{Reid} and {Hawley}(2005)}]{2005nlds.book.....R}
{Reid} IN {Hawley} SL (2005) {New light on dark stars : red dwarfs, low-mass
  stars, brown dwarfs}. \doi{10.1007/3-540-27610-6}

\bibitem[{{Reid} et~al.(2007){Reid}, {Cruz}, and {Allen}}]{2007AJ....133.2825R}
{Reid} IN, {Cruz} KL {Allen} PR (2007) {Meeting the Cool Neighbors. XI. Beyond
  the NLTT Catalog}. \aj 133:2825--2840

\bibitem[{{Reid} et~al.(2008){Reid}, {Cruz}, {Kirkpatrick}, {Allen}, {Mungall},
  {Liebert}, {Lowrance}, and {Sweet}}]{2008AJ....136.1290R}
{Reid} IN, {Cruz} KL, {Kirkpatrick} JD et~al. (2008) {Meeting the Cool
  Neighbors. X. Ultracool Dwarfs from the 2MASS All-Sky Data Release}. \aj
  136:1290--1311

\bibitem[{{Ricci} et~al.(2013){Ricci}, {Isella}, {Carpenter}, and
  {Testi}}]{2013ApJ...764L..27R}
{Ricci} L, {Isella} A, {Carpenter} JM {Testi} L (2013) {CARMA Interferometric
  Observations of 2MASS J044427+2512: The First Spatially Resolved Observations
  of Thermal Emission of a Brown Dwarf Disk}. \apjl 764:L27

\bibitem[{{Schmidt} et~al.(2014){Schmidt}, {Prieto}, {Stanek}, {Shappee},
  {Morrell}, {Bardalez Gagliuffi}, {Kochanek}, {Jencson}, {Holoien}, {Basu},
  {Beacom}, {Szczygie{\l}}, {Pojmanski}, {Brimacombe}, {Dubberley}, {Elphick},
  {Foale}, {Hawkins}, {Mullins}, {Rosing}, {Ross}, and
  {Walker}}]{2014ApJ...781L..24S}
{Schmidt} SJ, {Prieto} JL, {Stanek} KZ et~al. (2014) {Characterizing a Dramatic
  {$\Delta$}V \~{} -9 Flare on an Ultracool Dwarf Found by the ASAS-SN Survey}.
  \apjl 781:L24

\bibitem[{{Schneider} et~al.(2011){Schneider}, {Dedieu}, {Le Sidaner},
  {Savalle}, and {Zolotukhin}}]{2011A&A...532A..79S}
{Schneider} J, {Dedieu} C, {Le Sidaner} P, {Savalle} R {Zolotukhin} I (2011)
  {Defining and cataloging exoplanets: the exoplanet.eu database}. \aap 532:A79

\bibitem[{{Seager} et~al.(2009){Seager}, {Deming}, and
  {Valenti}}]{2009ASSP...10..123S}
{Seager} S, {Deming} D {Valenti} JA (2009) {Transiting Exoplanets with JWST}.
  Astrophysics and Space Science Proceedings 10:123

\bibitem[{{Seager} et~al.(2016){Seager}, {Bains}, and
  {Petkowski}}]{2016AsBio..16..465S}
{Seager} S, {Bains} W {Petkowski} JJ (2016) {Toward a List of Molecules as
  Potential Biosignature Gases for the Search for Life on Exoplanets and
  Applications to Terrestrial Biochemistry}. Astrobiology 16:465--485

\bibitem[{{Shvartzvald} et~al.(2017){Shvartzvald}, {Yee}, {Calchi Novati},
  {Gould}, {Lee}, {Beichman}, {Bryden}, {Carey}, {Gaudi}, {Henderson}, {Zhu},
  {Albrow}, {Cha}, {Chung}, {Han}, {Hwang}, {Jung}, {Kim}, {Kim}, {Kim}, {Lee},
  {Park}, {Pogge}, {Ryu}, and {Shin}}]{2017arXiv170308548S}
{Shvartzvald} Y, {Yee} JC, {Calchi Novati} S et~al. (2017) {An Earth-mass
  Planet in a 1-AU Orbit around a Brown Dwarf}. ArXiv e-prints

\bibitem[{{Terquem} and {Papaloizou}(2007)}]{2007ApJ...654.1110T}
{Terquem} C {Papaloizou} JCB (2007) {Migration and the Formation of Systems of
  Hot Super-Earths and Neptunes}. \apj 654:1110--1120

\bibitem[{{Wheatley} et~al.(2017){Wheatley}, {Louden}, {Bourrier},
  {Ehrenreich}, and {Gillon}}]{2017MNRAS.465L..74W}
{Wheatley} PJ, {Louden} T, {Bourrier} V, {Ehrenreich} D {Gillon} M (2017)
  {Strong XUV irradiation of the Earth-sized exoplanets orbiting the ultracool
  dwarf TRAPPIST-1}. \mnras 465:L74--L78

\bibitem[{{Whelan} et~al.(2005){Whelan}, {Ray}, {Bacciotti}, {Natta}, {Testi},
  and {Randich}}]{2005Natur.435..652W}
{Whelan} ET, {Ray} TP, {Bacciotti} F et~al. (2005) {A resolved outflow of
  matter from a brown dwarf}. \nat 435:652--654

\bibitem[{{Winn}(2010)}]{2010arXiv1001.2010W}
{Winn} JN (2010) {Transits and Occultations}. ArXiv e-prints

\bibitem[{{Wolszczan} and {Frail}(1992)}]{1992Natur.355..145W}
{Wolszczan} A {Frail} DA (1992) {A planetary system around the millisecond
  pulsar PSR1257 + 12}. \nat 355:145--147

\bibitem[{{Zeng} et~al.(2016){Zeng}, {Sasselov}, and
  {Jacobsen}}]{2016ApJ...819..127Z}
{Zeng} L, {Sasselov} DD {Jacobsen} SB (2016) {Mass-Radius Relation for Rocky
  Planets Based on PREM}. \apj 819:127

\end{thebibliography}

\end{document}